\RequirePackage{fix-cm}
\documentclass[twocolumn]{svjour3}          
\usepackage[english]{babel}
\usepackage{amssymb}
\newtheorem{ansatz}{Ansatz}
\usepackage{graphicx}

\begin{document}

\title{Maxwell-like picture of General Relativity and its Planck limit}
\subtitle{}


\author{Piotr Ogonowski         \and
        Piotr Skindzier
}

\institute{\and Leon Kozminski Academy,  ul. Jagiellonska 57, 03-301 Warsaw, Poland 
              \email{piotrogonowski.pl@gmail.com}     \\      
           \and
           WFAiS Jagiellonian University, ul. Reymonta 4, 30-059 Krakow, Poland 
              \email{piotr.skindzier@uj.edu.pl}
}


\maketitle

\begin{abstract}
We show that Geroch decomposition leads us to Maxwell-like representation of gravity in $(3+1)$ metrics decomposition that may be perceived as Lorentz invariant version of GEM. For such decomposition we derive four-potential $V^\mu$ and gravitational field tensor $F^{\mu\nu}$ that is associated with gravitational interaction. Next we show that gravitational four-current $J^\mu$ derived for introduced four-potential produce energy-stress tensor and reproduce main General Relativity formula. Next we introduce valid Lagrangian and equations of motion that explains obtained results. At the end we introduce new approach to quantization of gravity that results in proper quantum values and is open to further generalization.
\end{abstract}

\section{Introduction}

Classical theories of gravity can be sorted according to their degree of generality and completeness of the approach: since Newton gravity, through Newton-Cartan geometric approach up to General Relativity. Attempts to quantize gravity are taken over last 90 years for all existing descriptions of gravity and has not yet reached a satisfactory outcome. \cite{journal-1}\cite{journal-2}. \\

In the article, we propose a new, Lorentz invariant GEM-like approach to the description of gravity, which should be understood as an intermediate link between Newton-Cartan description and the General Relativity. The main reason to introduce this approach is that in this framework we may propose a method for efficient quantization of the gravity.

Our approach is derived from special case of General Relativity (Schwarzschild case) and shows, that in this case curved spacetime is physically equivalent to flat manifold minimally coupled to the scalar field. We also show, that in introduced framework we may generalize main GR equation for the field associated with gravitation.\\\\
We will call introduced approach "Dilation as Field" (DaF) because of its specificity - derived here scalar field is equal to the inversed gravitational time dilation factor. 
~\\
\includegraphics[scale=0.35]{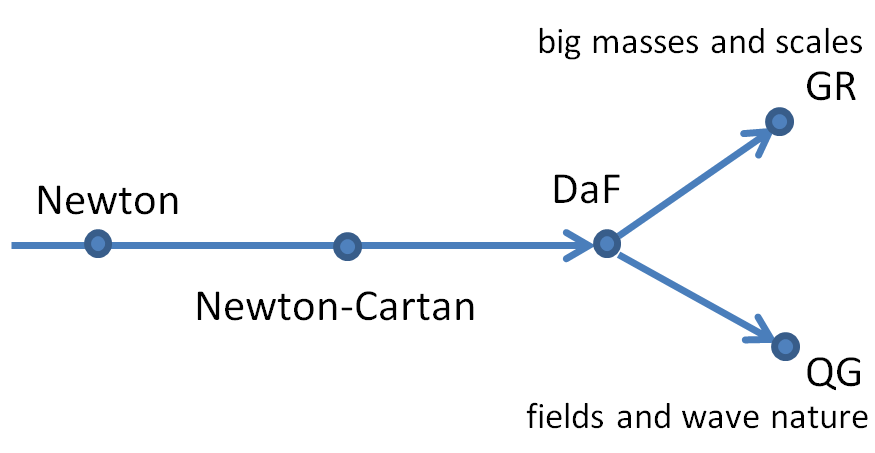}
~\\
Of course, there are known scalar theories of gravity on flat spacetime \cite{journal-3} \cite{journal-4} \cite{journal-5} and in most of these approaches authors try to extend gravity to the Planck limit. For these efforts the main problem appears to be generalization and effective quantization of these theories \cite{journal-6} \cite{journal-7}. In DaF we propose first to decompose General Relativity to (3+1) manifold, next generalize it in Maxwell-like picture and next consider obtained field in the Planck limits making quantization in classical and covariant way.
~\\
The paper is organized as follows.\\ 

In first part we show, that Geroch decomposition for Schwarzschild case leads us to Maxwell-like representation of gravity in $(3+1)$ metrics decomposition. In this decomposition we show, that gravitational action of a static spacetime is equivalent to the action of a 3-dimensional spacelike manifold minimally coupled to a scalar field $\Phi$. We also show, that gravity in such description may be perceived as the medium that changes the speed of light for chosen reference frame.\\

Next we propose description of that case with Maxwell-like equations on flat spacetime. In this perspective: 
\begin{itemize}
\item gravity is described by effective scalar field $\Phi$ equal to gravitational time dilation factor $\Phi=1/\gamma_r$ 
\item curved spacetime is physically equivalent to the flat spacetime with refracting index for light speed equal to $\gamma_r$ 
\end{itemize}
We also generalize obtained results and derive gravitational four-potential, four-current and field tensor. Then we show, that we can build energy-stress tensor thanks to introduced four-current for gravity and then we generalize main General Relativity formula for introduced field.\\

In third part of the article we show how we may quantize the field $\Phi$ to obtain proper rest mass and photon energy.
We also show, that parametrized $\Phi$ might serve us to describe classical electromagnetic field. We show that in infinity limit we are able to reconstruct electromagnetic equations in classic and covariant form and derive Coulomb-like potential for elementary charges.\\

We hope that this new approach to description gravity may bring important implications for our understanding of spacetime in zero limit, may shed new light on quantum gravity theories and opens new areas for research and generalizations.\\

\section{Geroch decomposition}
\subsection{Introducing Killing vector fields}

We will use Einstein summation convention: commas denote partial derivatives and semicolons denote covariant derivative. We choose metric signature $(+,-,-,-)$.\\

On a spacetime, a Killing vector field \cite{journal-8} generates an isometry of spacetime. If we have a coordinate chart in which the metric coefficients are independent of a coordinate, then the vector field of that coordinate is automatically a Killing field.\\

To introduce Killing fields we need to choose metric for which we can find at last one vector $\xi _{\mu }$. For the simplicity let us introduce Schwarzschild spacetime:
\begin{equation} ds^{ 2}=\left( 1-\frac{r_{s} }{r} \right) dt^{ 2}-\frac{dr^{ 2}}{\left( 1-\frac{r_{s} }{r} \right) } -r^{ 2}\left( d\theta ^{ 2}+\sin ^{ 2}\theta \,d\varphi ^{ 2}\right) \label{eq:schwcl}\end{equation}
We see that the metric coefficient $\left( 1-\frac{r_{s} }{r} \right)$ is time t and $\varphi  $ independent, so that $\partial _{t}  $ and $\partial _{\varphi }  $ are Killing fields.\\

\subsection{The metric and its interpretation}

Now, let us apply Geroch decomposition \cite{journal-9} for chosen spacetime to obtain requested metric description. To provide ease of interpretation we do not use ADM formalism \cite{journal-10}, however it might be useful for future generalizations.\\

For asymptotically flat spacetime with a timelike Killing vector field $\xi$ with norm-squared 
\begin{equation} \lambda =-\xi ^{ \mu} \xi _{\mu } \end{equation}
and twist 
\begin{equation} \omega _{\mu } =\epsilon _{\mu \nu \rho \delta } \xi ^{ \nu } \nabla ^{ \rho }\xi^{\delta} \end{equation}
using the tensor 
\begin{equation} \gamma _{\mu \nu } =\lambda g_{\mu \nu } +\xi _{\mu } \xi _{\nu } \end{equation}
the spacetime metric takes the form:
\begin{equation} ds^{ 2}=-\lambda \left( dt-\omega _{i} dx^{ i}\right) ^{ 2}+\frac{\gamma _{ij} }{\lambda } dx^{ i}dx^{ j} \end{equation}
Schwarzschild spacetime in the Schwarzschild coordinate chart $\xi = \partial _t$ gives:
\begin{equation}
\lambda = -\xi \cdot \xi = 1-r_s/r
\end{equation}
and the twist vanishes.\\
We may then rewrite metric to form of:
\begin{equation} ds^{2}=-\lambda dt^{2} + 2\lambda \omega _{i}dtdx^{i}-(\lambda \omega_{i} \omega_{j} +\frac{\gamma _{ij}}{\lambda} )dx^{i}dx^{j} \end{equation}
or easier:
\begin{equation}
 ds^{2}=g_{00} dt^{2} + 2g_{0i}dtdx^{i}+g_{\alpha \beta }dx^{i}dx^{j}   
\end{equation}

We have constructed a timelike Killing field what means that we have a stationary spacetime. If we normalize the four-velocity properly: 
\begin{equation} U^{ \mu} =\lambda ^{ -1/2}\xi^{\mu}  \end{equation}
we see, that
\begin{equation} \lambda ^{ -1/2} =\frac{1}{\sqrt{1-\frac{r_{s} }{r} } } =\gamma _{r} \label{eq:gammaGrav} \end{equation}
is the inverse-norm of timelike Killing field.\\

The four-acceleration of the Killing observers given by a covariant derivative 
in the static case simplifies to just: 
\begin{equation} a^{ \mu} =U^{ ;\mu} =g^{ \mu \nu} \left( \log \lambda ^{ 1/2}\right) _{,\nu } \end{equation}

Everything but the r-component vanishes, and we can put it into corresponding orthonormal basis rather then coordinate vector:
\begin{equation} \partial _{r}  \hat{e}_r=\partial _{r} \left(\sqrt{1-\frac{r_{s} }{r} } \partial _{r} \right) \end{equation}
wich yealds to:
\begin{equation} a^{ r}=\left( 1-\frac{r_{s} }{r} \right) \frac{1}{2} \frac{r_{s} }{r^{ 2}} \frac{1}{1-\frac{r_{s} }{r} } =\frac{r_{s} }{2r^{ 2}} \end{equation}  
thus
\begin{equation} a=\frac{r_{s} }{2r^{ 2}} \frac{1}{\sqrt{1-\frac{r_{s} }{r} } } \hat{e}_{r} =g \hat{e}_{r} \label{eq:przyspieszGrav} \end{equation}
showing the correct gravitational acceleration: the proper acceleration of the 
Killing observers is $g$ into the outwardly radial direction.\\

Of course this is only conceptual interpretation of our (3+1) decomposition of Schwarzschild metric. In this moment we have to stress that depending on coordinate system we can get different interpretation of gravity in pseudo classical picture.\\

However, above interpretation - the metric and the acceleration formula - are important pieces of information that will be requested in the section \ref{sec:Maxwell}.

\subsection{Action of the scalar field \label{sec:action}}

Let us show, that for chosen metric decomposition we may consider gravity through the scalar field.\\
For convention let us introduce:
\begin{eqnarray}
h&\equiv& g_{00}=-\lambda \\
g_{\alpha }&\equiv& -\frac{g_{0\alpha }}{g_{00}}=-2\lambda \omega _{i}\\
\Upsilon _{\alpha \beta }&\equiv& -g_{\alpha \beta }+hg_{\alpha }g_{\beta }=-( \frac{\gamma _{ij}}{\lambda} + \lambda \omega_{i} \omega_{j})
\end{eqnarray}

Taking t=0 hypersurface, the symmetry of the spacetime under time inversion means that the extrinsic curvature is zero, in which case the Gauss-Codazzi equations simplify to: 
\begin{equation} R^{ a}_{bcd} =^{h}R^{ a}_{bcd} \end{equation}
\begin{equation}  R^{ 0}_{bcd} =0 \end{equation} 
\begin{equation} R^{ 0}_{i0j} =-^{h}U_{;i;j} -\left( ^{h}U_{;i} \right) \left( ^{h}U_{;j} \right)  \end{equation}  
where the superscript h denotes that the quantity belongs to spatial hyperslice 
and should be calculated using the spatial metric h alone. \\

Contracting all the way down to the Ricci scalar:
\begin{equation}R=^{h}R-2\left[ ^{h}U^{ ;i}_{;i} +\left( ^{h}U^{ ;i}\right) \left( ^{h}U_{;i} \right) \right]  \end{equation}
The first term in the bracket is a Laplacian
\begin{equation}2U=\log \lambda \end{equation}
and the metric determinant is
\begin{equation}g=-\lambda h=-\gamma \end{equation}
so in terms of the metric: 
\begin{equation}\gamma _{ij} \left( \frac{1}{\lambda } \right) h_{ij} \end{equation}
the representation of the Einstein-Hilbert action is 
\begin{equation}S\;\alpha \;\int R\sqrt{-g}  \end{equation}
\begin{equation} S=\int \left(^{ \gamma} R-^{\gamma} \nabla ^{ 2}\left( \log \lambda \right) -\frac{1}{2} \frac{\gamma _{ij} d\lambda ^{ i}d\lambda ^{ j}}{\lambda ^{ 2}} \right) \sqrt{\gamma } \label{eq:action}\end{equation}  

We have obtained proper result. In this picture the gravitational action of a static spacetime is equivalent to the action of a 3-dimensional spacelike manifold minimally coupled to a scalar field $\Phi=1/\gamma_r$. \\

We will use above conclusion to introduce Maxwell-like equations for gravity on flat spacetime with $\Phi$ as scalar field in the section \ref{sec:Maxwell}.

\subsection{Maxwell-like equations considered locally}

Finally, we will show, that in chosen decomposition we may consider locally Maxwell-like equations, referring to the local time of stationary observer (Killing observer).\\
We may introduce Maxwell-like equations in (3+1) form:
\begin{equation}
    F_{\left [ ik;l \right ]}=0
\end{equation}

In more classical way:
\begin{equation}
\triangledown\cdotp \mathbf{B}=0
\end{equation}
\begin{equation}
\triangledown \times \mathbf{E} = -\frac{1}{\Upsilon }\frac{\partial }{\partial t}\left ( \sqrt{\Upsilon }\mathbf{B} \right )
\end{equation}
For better understanding let us write:
\begin{equation}
    \frac{1}{\sqrt{-g}}\partial_{a}\left ( \sqrt{-g}F^{ab} \right )=0
\end{equation}
then Maxwell-like equations can be rewritten as:
\begin{equation}
    \triangledown  \cdot \mathbf{D}=0
\end{equation}
\begin{equation}
\triangledown \times \mathbf{H} = -\frac{1}{\Upsilon }\frac{\partial }{\partial t}\left ( \sqrt{\Upsilon }\mathbf{D} \right )\\
\end{equation}

where:
\begin{equation}
    \mathbf{D}=\frac{\mathbf{E}}{\sqrt{h}}+\mathbf{H}\times \mathbf{g}
\end{equation}
\begin{equation}
 \mathbf{B}=\frac{\mathbf{H}}{\sqrt{h}}+\mathbf{g}\times \mathbf{E}
\end{equation}
Schwarzschild metric is static and components of it are time independent so we can rewrite Maxwell-like equations as:
\begin{eqnarray}
    \triangledown  \cdot \mathbf{B}=0\\ 
\triangledown \times \mathbf{E} = -\frac{\partial \mathbf{B}}{\partial t}\\
\triangledown  \cdot \left ( \frac{\mathbf{E}}{\sqrt{h}} \right )=0\\
\triangledown \times \mathbf{\sqrt{h}B} = \frac{1}{\sqrt{h}}\frac{\partial \mathbf{E}}{\partial t}\\
\end{eqnarray}
Above equations acts like Maxwell equations, where the speed of wave propagation is
\begin{equation} V_{w}=\sqrt{h}=c\sqrt{ 1-\frac{r_{s} }{r}} \label{eq:EqCspeed} \end{equation} 
It is simple in this moment to show that 
\begin{equation} \lim_{r \to \infty }V_{w}=c \end{equation} 

We may also generate Maxwell-like equations considering properties of the Killing fields.  \\
As we know Killing vector X\textsuperscript{b} by definition satisfies: 
\begin{equation}g^{ bc}X_{c;ab} -R_{ab} X^{ b}=0 \mbox{\cite[p. 443; C.3.9]{journal-11} } \end{equation}
\begin{equation} X_{a;bc} =R_{abcd} X^{ d} \mbox{\cite[p. 443; C.3.6]{journal-11} } \end{equation} 
\begin{equation} X^{ a;b}_{;b} +R_{c}^{a} X^{ c}=0 \end{equation} 
Thus for introduced Killing field we have: 
\begin{equation}R^{ \alpha} _{\beta \gamma \sigma } \xi ^{ \sigma} =\xi ^{ \sigma }_{;\beta ;\gamma } \end{equation}
Therefore:
\begin{equation}\xi _{\alpha ;\beta } ^{ ;\beta} =-R_{\alpha \beta } \xi ^{ \beta} \end{equation}
so the timelike Killing field is intimately connected to spacetime curvature. Defining a convenient F and using the defining property of Killing fields 
\begin{equation}\xi _{\alpha ;\beta } +\xi _{\beta ;\alpha } =0 \end{equation}
we obtain
\begin{equation}F_{\alpha \beta } =\xi _{\left[ \beta ;\alpha \right] } =\frac{1}{2} \left( \xi _{\beta ;\alpha } -\xi _{\alpha ;\beta } \right) =-\xi _{\alpha ;\beta } \end{equation}
therefore
\begin{equation}F_{\alpha \beta } ^{ ;\beta} =R_{\alpha \beta } \xi ^{ \beta} \end{equation}
This looks like the standard electromagnetic field tensor
\begin{equation}F_{\alpha \beta } ^{ ;\beta} =A_{\beta ;\alpha } -A_{\alpha ;\beta }  \end{equation}

In vacuum, the Ricci tensor vanishes, and the Killing field $\xi  $ seems to act like the electromagnetic four-potential A, that acts for electromagnetism in source-free regions, in the Lorentz gauge 
\begin{equation}A^{ \alpha} _{;\alpha } =0 \end{equation}
which automatically satisfies all Maxwell's equations.
Physical meaning of the tensor F should be explained as propagating spacetime anisometry.\\

Above Maxwell-like equations are known. In next sections we will show, there is a way to consider gravity through these equations.

\subsection{Conclusions}

We have just shown that in the Geroch decomposition for Schwarzschild metric we may obtain Maxwell-like picture of gravity that might let us to consider gravity as medium that changes the speed of light. Of course, in local coordinate system speed of light is still constant and equal to $c$ however in Geroch picture we clearly see equivalence between curved spacetime and flat spacetime with refracting index for light equal to $\gamma_r$.\\

This consideration leads us to the conclusion that after Geroch decomposition we can reconsider gravity not only as a spacetime curvature but also as some field  $\Phi=\sqrt{h}= 1/\gamma_r$. Of course, this picture of gravity we can only get in (3+1) Geroch decomposition and equation of motion of this field is dependent on what coordinate system we choose.\\

We have to be aware of this property of the gravity. If we try to interpret gravity as field we first need to decide what is the metric and in what coordinate system we consider this metric. In naive picture we can say that:

- metric choose is equivalent to choose of field (conservation laws),

- choose of coordinate system is equivalent to choose of equation of motion (dynamic equations).

\section{Maxwell-like representation of gravity (DaF) \label{sec:Maxwell}}

\subsection{Gravity as wave equation in Minkowski spacetime \label{sec:waveeq}} 

Let us suppose, that we would like to get Maxwell-like field tensor $F_{\mu \nu}$ for gravity expressed with some four-potential $V^{\mu}$ as
\begin{equation}
F_{\mu \nu} = \partial_\mu V_\nu- \partial_\nu V_\mu \label{eq:fieldTens}
\end{equation}
Of course, we could take the approach presented in GEM \cite{journal-GEM}, \cite{journal-GEM2} or other related approaches \cite{journal-ML}, \cite{journal-ML2}, \cite{journal-GML_Hol}, \cite{journal-GMInf}. 
However, we have shown in section \ref{sec:action} that scalar field $\Phi=1/\gamma_r$ may serve us to describe gravity on flat manifold. Therefore we will start with such scalar field and following  \cite{journal-12} we will consider some Maxwell-like equations on flat manifold.

It is known, that Special Relativity was invented as consequence of Maxwell rights analysis. Considering Liénard – Wiechert potential we may also find, that Maxwell equations are already relativistic. Therefore we will start with classical Maxwell-like definitions for gravity on flat manifold.\\
~\\
At first let us define scalar potential:
\begin{equation} \Phi=\frac{1}{\gamma _r}= \sqrt{1-\frac{r_s}{r}}=\frac{d\tau}{dt}\end{equation} 
where:
\begin{itemize} 
\item $t$ is coordinate time of the distant observer in infinity
\item $\tau$ is proper time of the local, stationary observer
\end{itemize}
~\\
Then we introduce vector field $\vec{G}$ as:
\begin{equation} \vec{G} = -\nabla \Phi \label{gradG} \end{equation}
After simple calculations we see, that its value (with respect to c=1 factor) appears equal to gravitational acceleration introduced in formula (\ref{eq:przyspieszGrav}).\\

Now we need to introduce some four-potential $V^\mu$ expressed with scalar $\Phi$ and some vector field $\vec{V}$ as
\begin{equation}
V^{\mu}= \left ( \Phi, \vec{V} \right )
\end{equation}
Therefore we should be able to express introduced vector field $\vec{G}$ responsible for gravitational acceleration in form of:
\begin{equation} \vec{G} = -\nabla \Phi - \frac{\partial \vec{V}}{\partial \tau} \label{eg:Lorgau} \end{equation}
and - by analogy to electromagnetism - we should be able to express $\vec{V}$ by some new field $\vec{\Omega}$ as:
\begin{equation} \vec{\Omega} = \nabla \times \vec{V}  \label{rotac} \end{equation} 

To obtain requested result we will define following vector fields (where $\hat{e}$ are directional versors):
\begin{equation}\vec{T} = \Phi \cdot \hat{e}=\vec{\frac{1}{\gamma _r}}  \label{eq:Tfield} \end{equation}
\begin{equation} \vec{G} = -\nabla \Phi \times \hat{e} = -\nabla \times \vec{T} \end{equation}
\begin{equation} \vec{\Omega } = \nabla \times \vec{V} = \nabla \beta_r \times \hat{e} \label{rotac} \end{equation} 
\begin{equation} \vec{V} = \vec{v_r} \end{equation}
This way we have obtained:
\begin{equation}
V^{\mu}= \left ( \Phi, \vec{V} \right )= \left ( \frac{1}{\gamma _r}, \vec{v_r} \right ) \label{fourpotent}
\end{equation}
As it is easy to check we obtain requested relations:
\begin{equation} \nabla \times \vec{\Omega} = -\frac{\partial \vec{G}}{\partial\tau} \end{equation}
\begin{equation} \nabla \times \vec{G} = \frac{\partial \vec{\Omega}}{\partial\tau} \label{Gprop} \end{equation} 
~\\
After simple transformations we derive d'Alembertian:
\begin{equation} \frac{\partial ^{ 2}\vec{G} }{ \partial \tau^{2}} - \nabla^2 \vec{G} = 0 \end{equation}
or with the same meaning:
\begin{equation}\gamma^2_r \cdot \frac{\partial ^{ 2}\vec{G} }{ \partial t^{2}} - \nabla^2 \vec{G} = 0 \label{eq:dAlem} \end{equation}
~\\
We have obtained wave with the same property as in (\ref{eq:EqCspeed}) formula. Above d'Alembertian, with respect to some factor, should be also able to work as electromagnetic wave description.\\
~\\
We know, that electromagnetic wave caries energy. In above description we see the wave that caries energy in terms of General Relativity (3+1) decomposition - it affects the metric the way we should expect. In section \ref{sec:quantum} we will show how we might associate above wave with the light wave.\\
~\\
Now, we may consider field tensor for Maxwell-like gravity. Introduced vector fields yields to field tensor for gravity equal to:
\begin{equation}
F_{\mu \nu} = \pmatrix{0 & G^1 & G^2 & G^3 \cr -G^1 & 0 & -\Omega^3 & \Omega^2 \cr -G^2 & \Omega^3 & 0 & -\Omega^1 \cr -G^3 & -\Omega^2 & \Omega^1 & 0 \cr}
\end{equation}
Above tensor should serve us gravitational field description at least for Schwarzschild case.\\\\
By analogy to four-current in electromagnetism we may also introduce four-vector:
\begin{equation}
J^{\mu} =\partial^\nu \partial_\nu V^\mu=\partial_\nu F^{\mu \nu} \label{eq:forcur}
\end{equation}
After simple calculations we may express introduced four-current as:
\begin{equation}
J^\mu = \rho U^\mu(r)
\end{equation}
where $U^\mu(r)$ is the four-vector resembling four-velocity:
\begin{equation}
U^\mu(r) = (\gamma_r, \beta_r \gamma_r)
\end{equation}
and $\rho$ would be the density of some gravitational pseudo "charge"
\begin{equation}
\rho = \frac{2\pi r_s}{V}  \label{eq:rhograv}
\end{equation}
~\\
In the energy-stress tensor in GR we consider energy density. Let us then check relation between above "gravitational charge density" $\rho$ and the energy density $\rho_m$. For given Schwarzschild radius it should be proportional up to some $\kappa$ constant:
\begin{equation}
\rho=\kappa \cdot \rho_m \;\;\; \to \frac{2\pi r_s}{V} = \kappa \cdot \frac{mc^2}{V}
\end{equation}
We calculate, that:
\begin{equation} \kappa = \frac{4 \pi G}{c^4} \end{equation}  
since
\begin{equation}
mc^2 = \frac{c^4 r_s}{2G}
\end{equation}
~\\
We recognize in above constant $\kappa$ half of Einstein's constant $8 \pi G /c^4$. Therefore, physical interpretation of introduced gravitational four-current $J^\mu$ might help us to simplify Einstein's stress-energy tensor.\\\\
Let us introduce the gravitational current tensor in the form of:
\begin{equation}
J^{\mu\nu}=J^{\mu}\cdot U^{\nu}
\end{equation}
where $U^{\nu}$ is the four-velocity of the source of gravity.\\
~\\
We see, that $J^{\mu \nu}$ is proportional, up to parameter $\kappa$, to the Stress-Energy Tensor introduced in General Relativity:
\begin{equation}
J^{\mu\nu} = \kappa \cdot T^{\mu\nu}
\end{equation}
Therefore $J_{\mu\nu}$ is also proportional to the Einstein curvature tensor $G_{\mu\nu}$
\begin{equation}
G_{\mu\nu} = 2 \cdot J_{\mu\nu}  \label{eq_curv}
\end{equation}

The equation says, that spacetime curvature depends on the distribution of the introduced four-current for gravitational field. This way we have created General Relativity main equation equivalence as wave-based formulation that might help us with explaining the wave nature of mater.\\
~\\
We also see here, that starting with four-potential on flat manifold, we may easy switch to the description for curved spacetime. This possibility comes from our findings in section \ref{sec:action}, that flat manifold minimally coupled to the introduced scalar field may be perceived as curved spacetime.\\
~\\ 
However, we may still come back to the flat spacetime picture and consider Lagrangian and equations of motion, looking for more general formula than GR, referring directly to the gravitational field tensor.

\subsection{Lagrangian, Hamiltonian and equations of motion}

At first let us derive Lagrangian and Hamiltonian for the action related to the scalar field  $\Phi=1/\gamma_r$ on flat manifold.\\

Joining together conclusions (\ref{eq:przyspieszGrav}) and (\ref{gradG}) we may write for stationary, Killing observer, that his proper acceleration g (up to c=1 factor) is equal to the gradient of $\Phi=1/\gamma_r$ what in this case yields to:
\begin{equation}
g=\frac{d(v_r \gamma_r)}{d\tau}=\frac{d \left( \frac{1}{\gamma_r} \right ) }{dr} \label{eq:lag_like}
\end{equation} 
Let us see, that it looks like the standard Euler-Lagrange condition where we have reduced the mass.\\
Taking $ E_0=mc^2 $ let us define:
\begin{equation}
L= E_0 \cdot \frac{1}{\gamma_r}
\end{equation}
then, let us define radial move in form of:
\begin{itemize}
\item "$\cdot$" represents $\frac{d}{d\tau}$
\item $\dot{r}=v_r$
\item $\beta_r = \frac{v_r}{c}$
\end{itemize}
Considering only radial move we obtain Hamiltonian in form of:
\begin{equation}
H=\dot{r}\frac{\partial L}{\partial \dot{r}} - L = - E_0\beta_r^2\gamma_r - E_0 \cdot \frac{1}{\gamma_r} 
\end{equation}
\begin{equation}
H=- E_0\gamma_r \left ( \beta_r^2 + \frac{1}{\gamma_r^2} \right )=-E_0\gamma_r
\end{equation}
what - as expected - represents the amount of energy, that stationary Killing observer have to have to keep his position against gravity.\\
~\\
As we may see here, gravitational potential in our approach is equivalent to the move against gravitational force. We have already seen this move for Killing observer in (\ref{eq:przyspieszGrav}). Further generalization of the Lagrangian to the metrics other than Schwarzschild would need adding other than radial degrees of freedom.\\\\
Now, let us consider Lagrangian and Hamiltonian for test body.\\

For test body with rest energy $E_0$ moving in flat 3 dimentional space we would have Lagrangian in form of: 
\begin{equation} L= -E_0 \cdot \frac{1}{\gamma} \end{equation}
and Hamiltonian in form of:
\begin{equation} H=\sum_{i=1}^3 \dot{x}_i\frac{\partial L}{\partial \dot{x}_i} - L=E_0\gamma \end{equation}
where:
\begin{itemize}
\item $s$ is proper time of the test body
\item $\tau$ is the proper time of stationary observer
\item $\gamma = \frac{d\tau}{ds}$
\end{itemize}

Thanks to superposition principle we may introduce Lagrangian for the test body in gravitational potential in form of: 
\begin{equation} L= E_0 \left ( \frac {1}{\gamma_r} - \frac {1}{\gamma} \right ) \end{equation}
and Hamiltonian in form of:
\begin{equation}H= E_0 \gamma - E_0 \gamma_r \label{HamiltonSimple} \end{equation}
To comply with the Newton approximation we will note Hamiltonian in form of:
\begin{equation} H= E_0 (\gamma-1) - V (r) \label{Hamilton} \end{equation}
where
\begin{equation}
V(r)=E_0 (\gamma_r-1)  \label{eq:potential}
\end{equation}

The Newton's limit of gravitational potential we obtain by Maclaurin's expansion:
\begin{equation} V(r)=E_0 (\gamma_r -1) \approx mc^2\frac{\beta^2_r}{2} = m \frac{c^2 r_s}{2r} = G\frac{mM}{r} \end{equation}
~\\
As we may easy calculate, introduced Lagrangian locally satisfies Euler-Lagrange condition.
\begin{equation}
\frac{d(m\vec{v}\gamma)}{d\tau} = \frac{d(E_0\frac{1}{\gamma_r})}{dr}=mg \label{eq:solSh}
\end{equation}
~\\
It is worth to notice, that we would obtain the same results for the test body by adding some additional (e.g. time-like) coordinate $x_0=r$. By defining zero-dimension radial velocity $\dot{x_0}=v_r$ (representing fight against gravity) and considering Lagrangian and Hamiltonian for $(\mu=0,1,2,3)$ we obtain the same, valid results.\\\\
We suppose, that this remark might shed new light on time-like zero-dimension parameter that we consider in four-vectors.\\
~\\
Now, we will derive more general equations of motion and introduce more general formula related to the gravitational field tensor.   \\\\
From (\ref{eq:lag_like}) we have (assuming c=1).
\begin{equation}
\frac{d(\vec{v}\gamma)}{d\tau} = \nabla \Phi \label{eq:ELmot}
\end{equation}
For L and H that do not depend on time we may rewrite it as a four-vector:
\begin{equation}
\frac{d \left (\gamma, \vec{v}\gamma \right )}{d\tau} = \left ( \frac{\Phi}{d\tau}, \nabla \Phi \right )
\end{equation}
What we may multiply by $\gamma$ to obtain:
\begin{equation}
A^\mu = \left ( \frac{d\Phi}{ds}, \gamma \nabla \Phi \right ) \label{eq:acc_grad}
\end{equation}
Thanks to (\ref{eg:Lorgau}) in the Lorentz gauge it is just:
\begin{equation}
A^\mu =  \left ( \frac{d\Phi}{ds}, \frac{d\vec{v_r}}{ds} \right ) \label{eq:movSource}
\end{equation}
In above generalized equations of motion we also see, that the source of gravity is - just the move.\\
~\\
We could build more general acceleration tensor, that would measure acceleration also with respect to the spatial coordinates. Such tensor the same time represents the field in space:   
\begin{equation}
A^{\mu\nu} =\partial_\nu V^\mu = (\vec{G}+ i\vec{\Omega}) \label{eq:thefield}
\end{equation}
~\\
This way, we have obtained first element of the gravitational field tensor (\ref{eq:fieldTens}) where the second element is the acceleration tensor for the observers. If we consider it for a moment - it is exactly what we know from GR findings.\\\\
As it is easy to check, we may express Einstain curvature tensor by above gravitational field tensor as: 
\begin{equation}
G^{\mu\nu} = 8 \pi \left ( F^{\mu\alpha}F^\nu_\alpha-\frac{1}{4}g^{\mu\nu}F_{\alpha\beta}F^{\alpha\beta} \right )
\end{equation}

\subsection{Conclusions}

We have just seen, there exists Maxwell-like description of gravity. In this picture curved spacetime is physically equivalent to flat manifold coupled to the scalar field what affects light tracks making it curved. We have seen, there exist valid Lagrangian and Hamiltonian formulation for considered picture. We have also derived equations of motion and obtained very interesting generalization. \\
~\\
We should also point out, that thanks to antisymmetry of the introduced gravitational tensor we may also consider "repulsion of light tracks". As it is easy to check it results with spacetime expansion effect what could open new area for research for Derk Energy phenomena explanation. \\
~\\
We have also seen, that in presented DaF framework, gravitational tensor is explained as the consequence of the move. As we see from (\ref{eq:movSource}) any four-velocity $U^\mu=(\gamma,\gamma \vec{v})$ the same time is the source of gravitational four-potential $V^\mu=(1/\gamma,\vec{V})$\\
~\\
Now we should care about quantum picture of the introduced field. Therefore in next section we will check, if the field produce valid quantum values.

\section{Quantum picture of the field \label {sec:quantum}}

As it was shown, the interpretation of our problem will depend not only on metric we choose but also on coordinate system. This remark is our first condition if we would like to quantize classical field that we present in section \ref{sec:Maxwell}. \\

Up to now it is also not obvious how to choose Hamiltonian in a way that will give us equation of motion for operators and produce valid quantum values. We must also check if we can use Copenhagen interpretation of wave function.\\

We will test three approaches to test quantum picture of DaF framework and examine our results:\\\\
1) In first approach we will check the Planck limits of introduced in (\ref{eq:potential}) potential V(r)\\
2) In second approach we will write Hamiltonian in covariant way and choose condition for metric that will give us confidence that our equations of motion are well defined

\subsection{Planck limit of the potential V(r)}

It is worth to show, that calculating Planck limits of the introduced potential V(r) we obtain (after approximation with Maclaurin's expansion) valid classical approximation of requested quantum values.
\subsubsection {Rest mass as the field quantum value}

Assuming $r_s << l_{P}$
\begin{equation} \lim_{E_0 \to E_P; r \to l_P} V(r) = \end{equation}
\begin{equation}=E_{P} \left( \frac{1}{\sqrt{1-\frac{r_{s} }{l_P} } } -1\right) \approx m_{P} \cdot \frac{c^2 r_{s} }{2l_{P} } =\frac{c^{ 4}r_{s} }{2G}=mc^2 \label{eq:restMass} \end{equation}

We have obtained the value that may be treated as some rest energy (some rest mass M) related to given Schwarzschild radius.\\
~\\
Therefore potential energy in gravitational field we may consider as the Planck limit of the field in presence of the field:
\begin{equation} \lim_{E \to E_P} \left ( \lim_{r \to l_P} V(r) \right ) \cdot (\gamma_r-1) \approx mc^2 (\gamma_r-1) \approx G\frac{mM}{r}  \label{eq:finpot} \end{equation}

\subsubsection {Free photon energy \label{sec:ph}}

Considering rotating vector fields as in (\ref{rotac}) we might suppose that for rotation with "c" speed on radius "r" with "T" period we have: 
\begin{equation}
2\pi r = cT \label{eq:rotat}
\end{equation}
Let us calculate Planck limits for such situation: 

\begin{equation} \lim_{E_0 \to E_P; r_s \to l_P} V(r) = \end{equation}
\begin{equation}=E_{P}\left( \frac{1}{\sqrt{1-\frac{l_{P} }{r} } } -1\right) \approx \frac{\hbar}{2}  \frac{c}{r} = \frac{1}{2} \hbar \omega \label{eq:photonEnergy} \end{equation}
where we have from (\ref{eq:rotat}) that $\omega=c/r=2\pi / T$\\
~\\
As we see, considering twisted pair of above we obtain valid approximation for photon energy. We should notice, that above hypothetical photon energy formula may be tested for pulsations close to Planck pulsation $\omega_{P}=1/t_{P}$ treating Planck pulsation $\omega_{P}$ as the limit.\\

Considering in quantum mechanics a photon described by the commonly used energy formula $\hbar \omega$ we obtain known problems at Planck length scales that have vital meaning for quantum cosmology and for attempts to grand unification \cite{journal-13}.\\

Just introduced formula based on (\ref{eq:photonEnergy}) does not crash at Planck time scales, because $ \hbar \omega $ acts only as approximation for small energies. We will also confirm above findings in the next sections.

\subsubsection{Coulomb potential energy and fine structure constant approximation}

In \cite{journal-14} authors consider additional axis proposed in Kaluza-Klein theory as time-like imaginary axis, and they obtain valid electromagnetic field description. Let us follow conclusion from section \ref{sec:waveeq} and consider this imaginary axis as Euler helix based, mathematical interpretation of the rotation introduced in (\ref{rotac}) \\
~\\
As we know from (\ref{eq:restMass}) Planck limit of the V(r) for $r \to l_P$ has produced the rest mass. From (\ref{eq:rotat}) we know, that we may consider $cT=2\pi r$\\
Now, we will show, that for $$r_s \to l_P; r \to 2\pi l_P (=cT)$$ we obtain valid elementary charge and Coulomb potential approximation.
By analogy to (\ref{eq:finpot}) below we consider Planck limit of the field in presence of the field:
\begin{equation} \lim_{E \to E_P; r_s \to l_P} \left ( \lim_{r \to 2 \pi l_P} V(r) \right ) \cdot (\gamma_r-1) = \end{equation}
\begin{equation}= E_{P} \cdot \left( \frac{1}{\sqrt{1-\frac{l_{P} }{2\pi \,l_{P} } } } -1\right) \cdot 
\left( \frac{1}{\sqrt{1-\frac{l_{P} }{2\pi \,r } } } -1\right) \approx \end{equation} 

\begin{equation} \approx E_{P} \cdot 4\pi \alpha \cdot \frac{l_P}{4 \pi \,r} \approx \frac{\hbar c}{r} \cdot \alpha \label{eq:charge} \end{equation}

where $\alpha$ was put in place of obtained fine structure constant value approximation. This way we have obtained Coulomb potential energy expressed with natural units for two elementary charges. This shows that we may get in natural way fine structure constant approximation if we introduce interaction between two particles and consider it in first order approximation.

\subsubsection {Lorentz force transformation \label{sec:Lorentz}}

Thanks to (\ref{eq:charge}) we may assign to elementary charge, some value in energy units:
\begin{equation}
q \equiv \lim_{E \to E_P; r_s \to l_P; r \to 2\pi l_P} V(r) \approx 4\pi  \alpha E_{P}
\end{equation}
Thanks to above property we could rewrite electromagnetic field tensor and transform Lorentz force. \\\\
For the body with rest energy "m" and elementary charge "q" (expressed with energy units thanks to introduced field definitions) we might rewrite Lorentz force (assuming c=1) in the form of:
\begin{equation}
m A^{\mu} = q U_\nu F_{(E)}^{\mu\nu}
\end{equation}
where we used (E) to denote rewritten Electromagnetic field tensor. Above we may shorten as:
\begin{equation}
k A^{\mu} = U_\nu F_{(E)}^{\mu\nu}
\end{equation}
where k is some constant equal to $k = m/q$.\\\\
Now, thanks to (\ref{eq:movSource}) we may express acceleration by Gravitational potential denoted with index (G):
\begin{equation}
k \frac{\partial V_{(G)}^{\mu}}{\partial \tau} = U_\nu F_{(E)}^{\mu\nu}
\end{equation}
Since k is constant, it just increases the value of the potential, and we may rewrite formula for increased potential as:
\begin{equation}
\frac{\partial V_{(G)}^{\mu}}{\partial \tau} = U_\nu F_{(E)}^{\mu\nu}
\end{equation}
We see here, that the move in electromagnetic field is the source of changes in the gravitational potential.
~\\
By guess we may postulate twin equation: 
\begin{equation}
\frac{\partial V_{(E)}^{\mu}}{\partial \tau} = U_\nu F_{(G)}^{\mu\nu}
\end{equation} 
where $V_{(E)}^{\mu}$ is for electromagnetic four-potential.\\
~\\
Above equation says, that changes of electromagnetic field potential (e.g. traveling light waves) determine the changes in the distribution of the energy. Since photons carries energy - we may expect it is truth.  However this equation needs experimental confirmation.

\subsection{Covariant quantization}
The last thing we would like to know is, if in our picture of the fields we may obtain Hamiltonian from which we could get quantum equations that will be easily interpreted by use of the Copenhagen \cite{journal-15} interpretation of wave function. This remark leads us to last approach to quantization. \\

In this approach we have to notice that natural interpretation of wave function is easy for Cartesian and Minkowski metrics. For both of this spaces we have well defined time and space variable which are separated and do not have singularities which are artifacts of coordinate system. For that metrics wave function is well defined. \\

If we try to introduce quantization for other metric we start with problem of good definition of proper time, separation between space and time variables and singularities. When we try to quantize metric of the form of equation (\ref{eq:schwcl}) we have good definition of time, separation between space and time variables, but as we easily can see  we have singularity for $r=r_s$ which is not the source of the field. \\

This problem is well known from at least 100 years and was solved by Eddington in 1924 \cite{journal-16}. He had proposed transformation to the isotropic coordinate using:
\begin{eqnarray}
r &=& r_1 {\left( 1 + \frac{G M}{2 c^2 r_1} \right)}^{2}\\
r_1 & =& \frac{r}{2}-\frac{G M}{2c^2}+\sqrt{\frac{r}{4}\left(r-\frac{2G M}{c^2}\right)}
\end{eqnarray}
and metric takes the form:
\begin{equation}
 {d s}^{2} = \frac{(1-\frac{GM}{2c^2 r_1})^{2}}{(1+\frac{GM}{2c^2 r_1})^{2}} \, c^2 {d t}^2 - \left(1+\frac{GM}{2c^2 r_1}\right)^{4}(dx^2+dy^2+dz^2)
\end{equation}
As it can be shown, from this coordinate metric we have only one singularity in $r=0$ which is physical and is source of the field.\\

Now, we would like to have some generic method to quantize metrics that have easy physical interpretation.

\begin{ansatz}
We can quantize gravity if the metric fulfill all the properties:
\begin{itemize}
\item time have proper local interpretation
\item $g_{ij} = 0 \mbox{ for all } i \ne j, \mbox{ where } i, j = 0,..,3$
\item singularities are only point source of the field
\end{itemize}
\end{ansatz}

For that choose of the metric we can take Hamiltonian in covaraint form \cite{journal-17}
\begin{equation}
H= \frac{1}{2}  g^{\alpha \beta }p_{\alpha } p_{\beta  }
\end{equation}
and quantize it to the form of 
\begin{equation}
H= \frac{1}{2} g^{\alpha \beta }\widehat{p_{\alpha }} \widehat{p_{\beta  }}
\end{equation}
We choose to take representation of four-momentum in form of:
\begin{equation}
\widehat{p_{\alpha }}=(\hat{E},\bold{\hat p}) = (i \hbar\frac{\partial}{\partial t}, -i \hbar\frac{\partial}{\partial \bold{r}})
\end{equation}

This leads us to the commutation rule:
$$\{ p^{\mu},p_{\nu}\}=const. \delta^{\mu}_{\nu}$$
For that we can define equation of the form:
\begin{equation}
    \frac{D p^{\xi}}{dt}=\{H, p^{\xi}\} + \frac{\partial p^{\xi}}{\partial t}
\end{equation}
That equation by definition we can rewrite to more simple form:
$$\frac{D p^{\xi}}{dt}-\frac{\partial p^{\xi}}{\partial t}=\{H, p^{\xi}\}$$
and simplifying we get:
\begin{equation}
    \Gamma ^{ \mu} _{\xi 0}p_{\mu}+ \{H, p^{\xi}\} p^{\xi}=0
\end{equation}
Which we can write in quantum form as:
\begin{equation}
    \Gamma ^{ \mu} _{\xi 0} \widehat{p_{\mu}} \Psi+ \{H, \widehat{p^{\xi}}\} \widehat{ p^{\xi}} \Psi=0
\end{equation}
This give us:
\begin{equation}
    \Gamma ^{ \mu} _{\xi 0} \widehat{p_{\mu}} \Psi+ c \widehat{p^{\xi}} \widehat{ p^{\xi}} \Psi=0
\end{equation}

For our metric we get two equations:
\begin{eqnarray}
 \Gamma ^{0} _{1 0} \widehat{p_{0}} \Psi+ c \widehat{ p^{1}}^2 \Psi&=&0\nonumber\\
\Gamma ^{1} _{0 0} \widehat{p_{1}} \Psi+ c \widehat{p^{0}}^2 \Psi&=&0\nonumber
\end{eqnarray}
which we rewrite in the form of:
\begin{eqnarray}
 \Gamma ^{0} _{1 0}g_{00} \widehat{p^{0}} \Psi+ c \widehat{ p^{1}}^2 \Psi&=&0\nonumber\\
\Gamma ^{1} _{0 0}g_{11} \widehat{p^{1}} \Psi+ c \widehat{p^{0}}^2 \Psi&=&0\nonumber
\end{eqnarray}
Here we see that $\Gamma ^{0} _{1 0}g_{00} = - \Gamma ^{1} _{0 0}g_{11}$ and if we add this two equations we get:
\begin{equation}
\widehat{ p^{1}}^2 \Psi + \widehat{p^{0}}^2 \Psi = 0
\end{equation}
This is the plain wave equation. In this picture gravity can be seen as free wave function of particle with no mass. \\

In our opinion only external field or interaction between two massive particle can create in this picture effective mass of particle that have gravity mass in classical picture. \\

This should not surprise us. We consider our field locally and start quantize it in local neighborhood around singularity. In this case mass that we consider is in fact 0. In our definition of metric proper choose of mass is Komar Mass \cite{journal-18} which for vacuum solution of Einstein equation is always zero if we integrate on volume different than infinity. \\

This remark in natural way leads us to interpretation that our quantization is proper for this choose of metric. Our solution for quantum equation should be plain wave function with no mass for local frame.

\section{Conclusions and open issues}

In this article we try to shed new light on our understanding of gravity. We have shown that gravity can be seen as classical Maxwell-like field that propagates in the vacuum as wave. We have also shown, that we can reassemble General Relativity main formula for introduced gravitational field what opens possibility to investigate the wave nature of matter in General Relativity.\\

We have shown, that considering Planck limits for introduced potential, we obtain valid quantum values of rest mass, photon energy, Coulomb potential energy and fine structure constant approximation. We have also pointed the relation between introduced gravitational field tensor and quantum wave function.\\

We have started our framework with the Schwarzschild metric as the special case of GR, that at first sight seems the worst starting point for generalization. However, it is well known, that quantum mechanics works only for chosen spacetimes - not for general case.\\

From \ref{sec:waveeq} we know, how to build the tensor for dislocating point source of gravitational field. As it was shown in the last section of the article, we may also transform Schwarzschild metric to the isotropic coordinates. With such "point sources of gravitational field" we are able to build energy-stress tensor for "the dust" build of set of moving Schwarzschild sources. This way we can reconstruct any gravitational system and create energy-stress tensor the same way that it was done in the original Einstein's reasoning.\\ 

Of course, it is extremely complicated for big gravitational systems. It would be much easier to benefit from coincidence, that: $$lim_{r \to l_P}V(r) \approx m_P\frac{r_s}{2l_P}$$ and call it "the rest mass".\\

Thanks to above, we could consider mass density instead the field. And it is exactly what we see in General Relativity from DaF framework perspective. \\

In DaF picture of gravity, Einstein went too far. GR uses idea of the "mass", where indeed there is no mass at all, but only gravitational field. For Planck scales we can not consider the mass anymore, so GR fails on the Planck scales.\\

In the introduction we proposed to place DaF framework between Newton-Cartan aproach and GR. This way we make one step back, to be able to make few steps forward. We left GR as good approximation for big masses and big scales, but for Quantum Gravity (QG) we have to turn right at the "DaF" signpost to keep "the field" and "wave nature of matter".\\

Above leads us to conclusion, that we may try to quantize gravitational field and introduce effective theory of quantum gravity. We may also conclude, that it is natural to consider gravity in (3+1) manifold. This choose gives us also opportunity to reintroduce time in quantum mechanics as regular dimension.\\

In covariant quantization we have shown that in local frame we see only massless particles traveling  with speed of light. This picture will change if we introduce interaction with other particles for example other photon. It requires further investigation but we can propose test of our model for high frequency resonator where we can test interactions between photons that wave functions have Planck scales and interact between each other. It would need designing the experiment, but it seems that we could also benefit from the research presented in \cite{experim}, \cite{experim2} and \cite{experim3}.

\section{Acknowlegments}
We would like to thank to prof. Claudio M. G. de Sousa for all his valuable remarks.


\begin{thebibliography}{0}

\bibitem{journal-1} Stephen Boughn, "Nonquantum Gravity", Found. Phys. (2009), Volume 39, Issue 4, pp 331-351
\bibitem{journal-2} Carlo Rovelli, "Notes for a brief history of quantum gravity", arXiv:gr-qc/0006061
\bibitem{journal-3}  Milutin Blagojevic, Gravitation and gauge symmetries, Institute of Physics Publishing, (2002) London
\bibitem{journal-4} Ye Xing-Hao, Lin Qiang, "Inhomogeneous Vacuum: An Alternative Interpretation of Curved Spacetime ", Chinese Phys. Lett. 25 (2008)
\bibitem{journal-5} M. Novello, E. Bittencourt, U. Moschella, E. Goulart, J. M. Salim, J. Toniato, "Geometric scalar theory of gravity",  arXiv:1212.0770
\bibitem{journal-6} J. Fernando Barbero G., Eduardo J. S. Villaseñor, "Quantization of Midisuperspace Models", arXiv:1010.1637
\bibitem{journal-7} A. Ashtekar, M. Pierri, "Probing Quantum Gravity Through Exactly Soluble Midi-Superspaces I", arXiv:gr-qc/9606085
\bibitem{journal-8} S. Weinberg, Killing Vectors.13.1 in Gravitation and Cosmology: Principles and Applications of the General Theory of Relativity. New York: Wiley, pp.~375-381, (1972)
\bibitem{journal-9}  R. Geroch, A. Held, and R. Penrose, A spacetime calculus based on pairs of null directions, J. Math. Phys. 14, 874 (1973)
\bibitem{journal-10} R. Arnowitt, S. Deser, C. W. Misner, "The Dynamics of General Relativity", arXiv:gr-qc/0405109
\bibitem{journal-11} Robert M. Wald, "General Realtivity", ISBN-13:978-0-226-87033-5
\bibitem{journal-GEM} Bahram Mashhoon, "Gravitoelectromagnetism: A Brief Review", arXiv:gr-qc/0311030v2
\bibitem{journal-GEM2} Qasem Exirifard, "GravitoMagnetic Field in Tensor-Vector-Scalar Theory", arXiv:1111.5210
\bibitem{journal-ML} Eduardo A. Notte-Cuello, Waldyr A. Rodrigues Jr, "A Maxwell Like Formulation of Gravitational Theory in Minkowski Spacetime", arXiv:math-ph/0608017
\bibitem{journal-ML2} Jeffrey D. Kaplan, David A. Nichols, Kip S. Thorne, "Post-Newtonian Approximation in Maxwell-Like Form", arXiv:0808.2510
\bibitem{journal-GML_Hol} Rong-Xin Miao, Jun Meng, Miao Li, "f(R) Gravity and Maxwell Equations from the Holographic Principle", arXiv:1102.1166
\bibitem{journal-GMInf} Federico Agustin Membiela, Mauricio Bellini, "Coupled inflaton and
electromagnetic fields from Gravitoelectromagnetic Inflation with Lorentz
and Feynman gauges", arXiv:1003.4175
\bibitem{journal-12} P. Ogonowski, Time Dilation as Field, Journal of Modern Physics, Vol. 3 No. 2, 2012, pp. 200-207
\bibitem{journal-13} James B. Hartle, "Quantum Cosmology: Problems for the 21st Century", arXiv:gr-qc/9701022
\bibitem{journal-14} J.Kocinski, M.Wierzbicki, "The Schwarzschild solution in a Kaluza-Klein theory with two times", arXiv:gr-qc/0110075
\bibitem{journal-15} Claus Kiefer, "On the interpretation of quantum theory - from Copenhagen to the present day", arXiv:quant-ph/0210152
\bibitem{journal-16} A S Eddington, The Mathematical Theory of Relativity, 2nd edition  (Cambridge University press), at sec. 43, p.93. (1924)
\bibitem{journal-17} B.A. Dubrovin, A.T. Fomenko, and S.P. Novikov, "Modern Geometry- Methods and Applications, Part I", Springer-Verlag, Berlin ISBN 0-387-90872-2 (1984)
\bibitem{journal-18} Misner, Thorne, Wheeler, "Gravitation", W H Freeman and Company. ISBN 0-7167-0344-0 (1973)
\bibitem{experim} Raymond Y. Chiao, Robert W. Haun, Nader A. Inan, Bong-Soo Kang, Luis A. Martinez, Stephen J. Minter, Gerardo A. Muñoz, Douglas A. Singleton, "A Gravitational Aharonov-Bohm Effect, and its Connection to Parametric Oscillators and Gravitational Radiation", arXiv:1301.4270
\bibitem{experim2} Shinsei Ryu1, Joel E. Moore1,2, Andreas W. W. Ludwig, "Electromagnetic and gravitational responses and anomalies in topological insulators and superconductors", Phys. Rev. B 85, 045104 (2012)
\bibitem{experim3} Michael Stone, "Gravitational anomalies and thermal Hall effect in topological insulators", Phys. Rev. B 85, 184503 (2012)
\bibitem{add-1} Kei-ichi Maeda, Misao Sasaki,* Takashi Nakamura* and Sho-ken Miyama, "A New Formalism of the Einstein Equations for Relativistic Rotating Systems", Prog. Theor. Phys. Vol. 63 No. 2 (1980) pp. 719-721
\bibitem{add-2} Claudio M. G. de Sousa, "On the frequency shift of gravitational waves", arXiv:gr-qc/0207052
\bibitem{add-3} Yaakov Friedman, Tzvi Scarr, "Covariant Uniform Acceleration", arXiv:gen-ph/1105:0492
\bibitem{add-4} L. A. Glinka, Grav. Cosmolo., 317 (2009)
 
\end{thebibliography}
\end{document}